\documentclass[10pt,twoside,dvips]{article}
\usepackage{amssymb}
\usepackage{amsmath}
\usepackage{multicol}
\usepackage[spanish,english]{babel}
\usepackage{graphicx}
\begin{document}

\title{{\Large Bead, Hoop, and Spring as a Classical Spontaneous Symmetry Breaking
Problem}}
\author{Fredy Ochoa$^{1}\thanks{%
e-mail: faochoap@unal.edu.co}$ and Jorge Clavijo$^{2}\thanks{%
e-mail: eclavijo@escuelaing.edu.co}$ \and $^{1}$Departamento de F\'{\i}sica,
Universidad Nacional de Colombia, \and $^{2}$Departamento de Ciencias B\'{a}%
sicas, Escuela Colombiana de Ingenier\'{\i}a. \\
Bogot\'{a}-Colombia}

\maketitle

\vspace{-0.7cm}
\begin{abstract}
We describe a simple mechanical system that involves Spontaneous Symmetry
Breaking. The system consists of two beads constrained to slide along a hoop
and attached each other through a spring. When the hoop rotates about a fixed
axis, the spring-beads system will change its equilibrium position as a
function of the angular velocity. The system shows two different regions of
symmetry separated by a critical point analogous to a second order
transition. The competitive balance between the rotational diynamics and the interaction of the spring causes an Spontaneous Symmetry
Breaking just as the balance between temperature and the spin interaction causes a transition in a ferromagnetic system. In addition, the
gravitational potential act as an external force that causes explicit
symmetry breaking and a feature of first-order transition. Near the transition point, the system exhibits a universal
critical behavior where the changes of the parameter of order is described
by the critical exponent $\beta =1/2$ and the susceptibility by $\gamma =1$.
We also found a chaotic behavior near the critical point. Through a
demostrative device we perform some qualitative observations that describe
important features of the system.
\end{abstract}


\begin{multicols}{2}

\section{Introduction}

Although gauge theories and Spontaneous Symmetry Breaking (SSB) \cite{Gauge} are rather abstract concepts which are associated with symmetries
operating on internal degrees of freedom, their significance and nature
emerges in a quite straightforward way through systems that exhibit
geometrical properties associated with space-time symetries, and that shares
the same basic features as models with gauge structure. In the
literature some works about analogies with classical
systems are found \cite{Bernstain}, where different mechanical models are used to illustrate
similar characteristics found in particle physics and solid state physics.
In this work with a simple mathematical treatment, we describe another
mechanical system that exhibits SSB and dynamical restitution of a discrete
symmetry in a similar fashion as a second order phase transition. In
contrast to the systems reported by the literature, our model is in a closer
relation with a thermodynamic system in the sense that the former use the
gravity as an essential external parameter to generate symmetry restitution, while ours may generate spontaneous breaking and restitution with only internal interactions, where the gravity act as an external field that causes explicit
symmetry breaking, such as external interactions cause explicit breaking in
thermal physics. The model shows not only interesting features that
resembles to the SSB in gauge theories, but also involves other related
concepts. For example, it can be used in a basic level of classical mechanics
as an application of rotational dynamics and to visualize the stability
of systems from energy diagrams. In a higher level, it can be introduced as
an example in the Lagrangian formulation of analytical mechanics. In a
thermodynamic framework, this model illustrates the behavior of a first- and second-order trasition according to the Tisza's definition. Futhermore, because of
the nature of the SSB, this system exhibits a critical behavior near the
transition point which belongs to the same universality class as a large
number of thermodynamic systems. Thus, this problem may provide instructive
insights that concern with more advanced topics into the classical study of
the critical phenomena in phase transitions. In addition, the system presents
interesting properties associated with nonlinear solutions, which may be
source of chaotic behavior and generates a simple application into the theory
of classical chaos.

\vspace{-0.5cm}

\section{Stability\label{Stability}}

\begin{figure}[t]
\centering\includegraphics[scale=0.4]{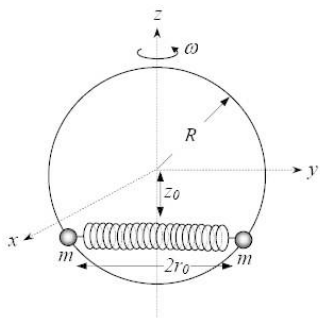}
\caption{{\footnotesize BHS model. Two beads attached by a spring slide without friction along a horizontal hoop
with rotation. The beads have the
same position $z_{0}$ from the center.}}
\label{draw}
\end{figure}

We consider two identical beads of mass $m$ that can slide without friction
on a horizontal hoop of radius $R$. There appears a potential of
interaction through a spring attached to both beads. The spring
constant is $k$ and its equilibrium length is less that the diameter of the
hoop, i.e. $2r_{0}<2R.$ The system is sketched in Fig. \ref{draw},
where the position $z_{0}$ of both beads is the same measured from the

\end{multicols}
\twocolumn

\noindent center of the hoop. There are two equivalent equilibrium positions determined
by the transformation $z_{0}\rightarrow -z_{0}$, which illustrates the
presence of a discrete symmetry. The hoop rotates about the $z$ axis at
angular velocity $\omega ,$ which causes the equilibrium position $z_{0}$ to
decrease and the spring to stretch from its natural length $2r_{0}$ to a
longer length $2r$. The lagrangian contains the kinetic energy of
each bead
and the elastic potential of the spring, while the gravitational potential is
not considered because the center of mass of the system lies along the
horizontal axis $z$.
It is more suitable to write the lagrangian in cylindrical
coordinates, which displays in a simpler form the symmetries and constraints
of the system. The lagrangian is then written as

\vspace{-0.3cm}

{\small \begin{equation}
L=m(\overset{\cdot }{r}^{2}+r^{2}\overset{\cdot }{\phi }^{2}+\overset{\cdot }%
{z}^{2})-2k(r-r_{0})^{2}.  \label{L-cil}
\end{equation}}

\vspace{-0.3cm}

The change in time of the azimuthal angle $%
\phi $ is identified as the angular velocity of the hoop. We can eliminate
the coordinate $r$ through the constraint $r=\pm \sqrt{R^{2}-z^{2}}$ which
indicates that the motion of the beads is restricted along the hoop. Hence,
the lagrangian can be expressed in terms of the generalized coordinate $z$ as

{\small \begin{equation}
L=\frac{1}{2}\mu \overset{\cdot }{z}^{2}-V_{ef}(z),  \label{L-gener}
\end{equation}}

where we have defined an effective mass $\mu =\frac{2mR^{2}}{R^{2}-z^{2}}$
and a one-dimensional effective potential given by

{\small \begin{equation}
V_{ef}(z)=V_{el}-V_{cent}=2k\left( \sqrt{R^{2}-z^{2}}-r_{0}\right)
^{2}-m\omega ^{2}\left( R^{2}-z^{2}\right) ,  \label{V-efec}
\end{equation}}

with $V_{el}$ the elastic potential of the spring and $V_{cent}$ the
centrifugal potential which contains the rotational dynamics. It is evident that the lagrangian in (\ref{L-gener}) does not change under the coordinate
transformation $z\rightarrow -z.$ The
introduction of the effective potential in Eq. (\ref{V-efec}) will help
us to determine the stability behavior of the system$.$ Equilibrium occurs
when

{\small \begin{equation}
\frac{dV_{ef}}{dz}=4kzr_{0}\left( \frac{1}{\sqrt{R^{2}-z^{2}}}-\frac{1}{\xi }\right)
=0,  \label{equil}
\end{equation}}

where $\xi (\omega )=\frac{2kr_{0}}{2k-m\omega ^{2}}.$ The condition (\ref%
{equil}) determines 3 equilibrium points at
\vspace{-0.05cm}
{\small \begin{equation}
z_{0}=0\quad \text{and} \quad \pm \sqrt{R^{2}-\xi ^{2}}.
\label{Z0}
\end{equation}}
\vspace{-0.2cm}

To investigate the stability of the solutions above, we must examine the
second derivative of the effective potential, which gives

{\small \begin{equation}
\frac{d^{2}V_{ef}}{dz^{2}}=4kr_{0}\left[ \frac{R^{2}}{\left(
R^{2}-z^{2}\right) ^{3/2}}-\frac{1}{\xi }\right] .  \label{stabil}
\end{equation}}

\begin{figure}[t]
\hspace{-2cm}\centering\includegraphics[scale=0.75]{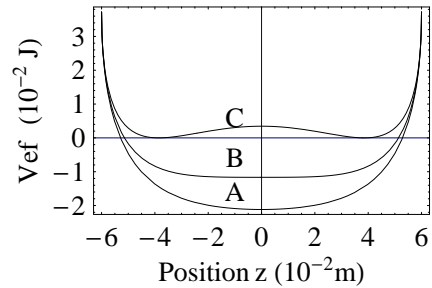}
\caption{{\footnotesize Effective potential as a function of the $z$ coordinate for angular
velocity (A) $\protect\omega >\protect\omega _{c},$ (B) $\protect\omega =%
\protect\omega _{c}$ and (C) $\protect\omega <\protect\omega _{c}$.}}
\label{symm-potential}
\end{figure}

This expresion evaluated in the 3 equilibrium points, displays different
possibilities according to the sign of the second derivative. The results
are resumed in table \ref{tab:stability}, where $R^{2}\geq \xi ^{2}$ was
demanded in order to assure real solutions for $z_{0}=\pm \sqrt{R^{2}-\xi ^{2}%
}.$ In Fig. \ref{symm-potential} we show three different plots of the
effective potential as a function of the coordinate $z$, where we
distinguish a critical angular velocity determined by

{\small \begin{equation}
\omega_{c}^{2}=\frac{2k(R-r_{0})}{mR},
\label{criticvel}
\end{equation}}

\begin{table}[b]
\begin{equation*}
\begin{tabular}{l|l|l}
\hline\hline
Angular Vel. & Equilib. points & Stability \\ \hline
$\omega ^{2}>\frac{2k(R-r_{0})}{mR}$ & $z_{0}=0$ & $\frac{d^{2}V_{ef}}{dz^{2}%
}>0$ {\small stab.}$$ \\ \hline
$\omega ^{2}=\frac{2k(R-r_{0})}{mR}$ & $z_{0}=0$ & $\frac{d^{2}V_{ef}}{dz^{2}%
}=0$ $${\small undef.}$$ \\ \hline
$\omega ^{2}<\frac{2k(R-r_{0})}{mR}$ & $%
\begin{tabular}{l}
\hspace{-0.2cm}$z_{0}=0$ \\ 
\hspace{-0.2cm}$z_{0}=\pm \sqrt{R^{2}-\xi ^{2}}$%
\end{tabular}%
$ & 
\begin{tabular}{l}
\hspace{-0.2cm}$\frac{d^{2}V_{ef}}{dz^{2}}<0$ $${\small unstb.}$$ \\ 
\hspace{-0.2cm}$\frac{d^{2}V_{ef}}{dz^{2}}>0$ $${\small stab.}$$%
\end{tabular}
\\ \hline\hline
\end{tabular}
\end{equation*}
\caption{{\footnotesize Stability regions according to the angular velocity. There is a
critical angular velocity that separates two different phases.}}
\label{tab:stability}
\end{table}

and two regions that depend on the value
of the angular velocity of the hoop. For values with $\omega ^{2}>$ $\omega
_{c}^{2}$ (symmetrical phase), there is just one minimum at $z=0$ that
corresponds to the case in which both beads hold their maxima distance in a
symmetrical position. When the angular velocity slowly decreases near to the
value determined by $\omega ^{2}=$ $\omega _{c}^{2},$ the minimum of energy
becomes flatter. Immediately below the critical
value $\omega _{c}$, the single minimum bifurcates into two degenerated
minima $z=\pm \sqrt{R^{2}-\xi ^{2}}$ with the same energy. However, the
spring chooses one of the minima causing an spontaneously symmetry breaking,
where the symmetry transformation $z\rightarrow -z$ is not manifest, and the
spring comes to a new equilibrium position in either a positive or a
negative coordinate $z$, which is greater with decreasing $\omega $. Thus, the
absolute value of $z$ measures the degree of breaking just like an order parameter.

\begin{figure}[t]
\hspace{-3.1cm} \includegraphics[scale=0.5]{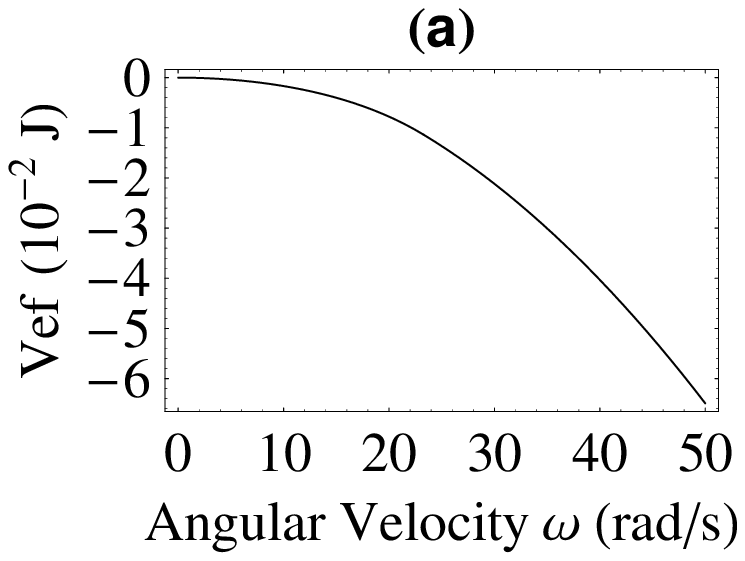} \hspace{-4.5cm} 
\includegraphics[scale=0.5]{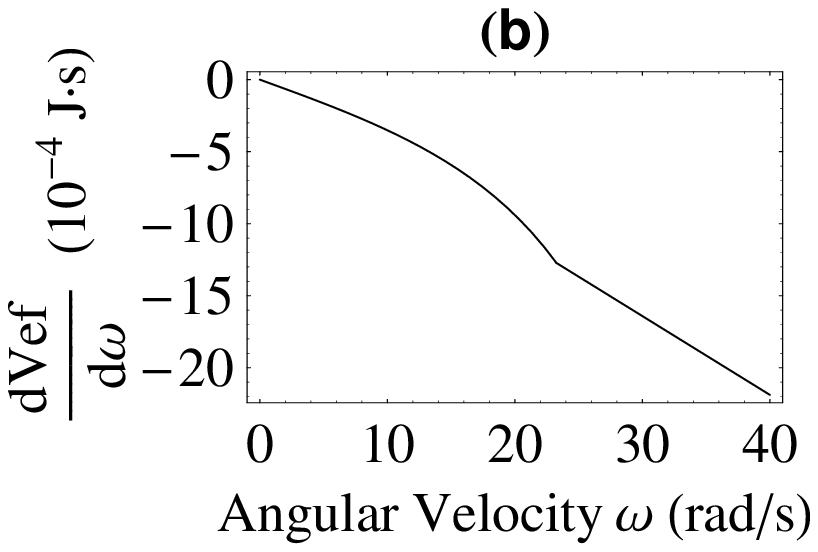} 
\includegraphics[scale=0.5]{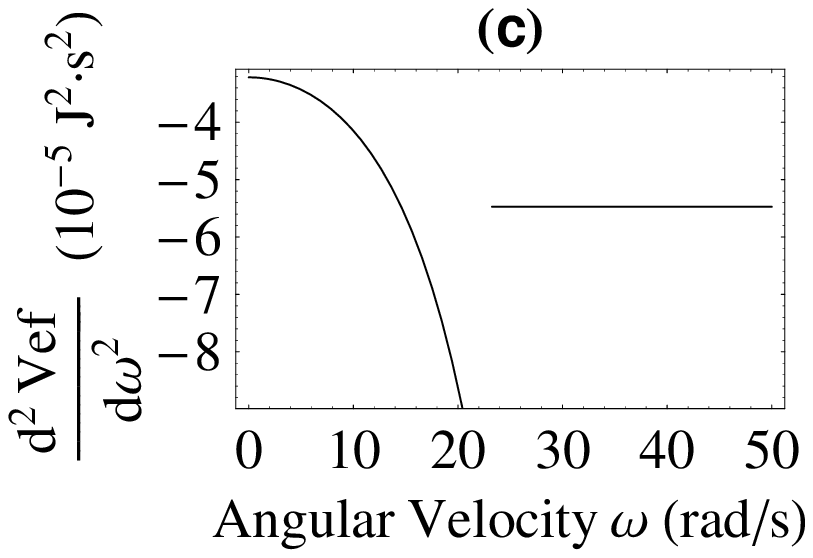}
\caption{{\footnotesize Analitycal behavior of the effective potential with the angular
velocity. (a) and (b) show the continuity of $V_{ef}$
and $\frac{dV_{ef}}{d\protect\omega }.$ (c) displays an
infinite discontinuity across the critical point $\protect\omega _{c}$ (in these plots is $23.26
rad/s$), exhibiting a second-order phase transition.}}
\label{Analitycal}
\end{figure}

We can describe this process like a second order phase transition if we
investigate the behavior of the effective potential in the critical point with
respect to the variation of the angular velocity. The Ehrenfest definition
establish that a phase transition with respect to a thermodynamic variable (like
temperature) is of $n^{th}$ order if the energy and its $\left( n-1\right)
st $-order derivatives are continuous at the transition point, whereas the $%
n^{th}$-order derivative suffers a finite discontinuous jump. In fact, our
system presents continuous derivative at first-order but suffers an infinite
discontinuity in the second-order derivative $\frac{d^{2}V_{ef}}{d\omega ^{2}%
}$ as is shown in Fig. \ref{Analitycal}$,$ where the angular velocity $%
\omega $ works like the temperature in a thermodynamic system. The infinite
jumps are described by the Tisza's Theory \cite{Callen} which describes most
of the higher-order phase transition.

We emphasize that this behavior is analogous but not equal to the SSB in
quantum theory. The analogy arises when the minimum of energy splits
into new degenerated minima. Nonetheless, classically the system will
maintain its symmetrical unstable equilibrium unless there appears external
fluctuations that are not being considered into the equation of motion (in
this sense this is not a "true" SSB). On the other hand, the quantum theory
introduces fluctuations associated with the uncertainty principle inherent
into the structure of the theory. Such fluctuations appear spontaneously to
unbalance the system toward a not symmetrical state .

If the horizontal hoop is turned into an angle $\alpha$, the gravity induces an
explicit symmetry breaking. In this case the effective potential in (\ref%
{V-efec}) changes as

\vspace{-0.3cm}
{\small \begin{equation}
V_{ef}(z)=2k\left( \sqrt{R^{2}-z^{2}}-r_{0}\right) ^{2}-m\omega ^{2}\left(
R^{2}-z^{2}\right)
+2mgz\text{Sin}\alpha,  \label{Non-sym}
\end{equation}}

\begin{figure}[t]
\hspace{-1cm}\includegraphics[scale=0.7]{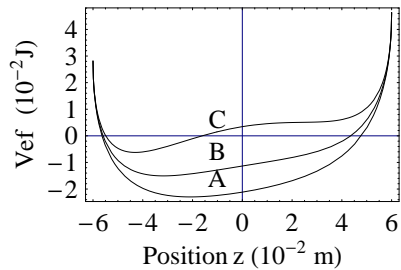}
\caption{{\footnotesize Effective potential as a function of the $z$ coordinate for angular
velocity (A) $\protect\omega >\protect\omega _{c},$ (B) $\protect\omega =%
\protect\omega _{c}$ and (C) $\protect\omega <\protect\omega _{c}$. The gravity causes the negative minimum to be
lower.}}
\label{nonsymm-potential}
\end{figure}

\begin{figure}[t]
\hspace{-1.8cm}\includegraphics[scale=0.48]{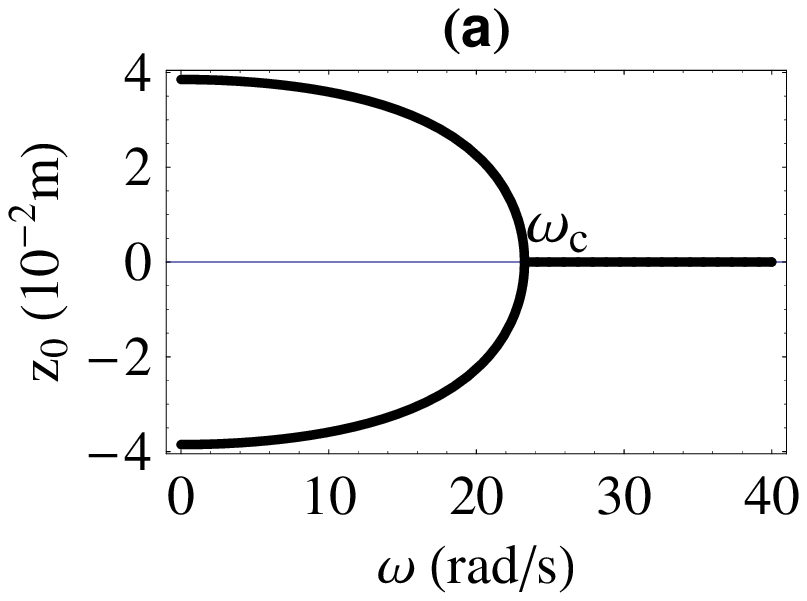} \hspace{-2.5cm}
\includegraphics[scale=0.49]{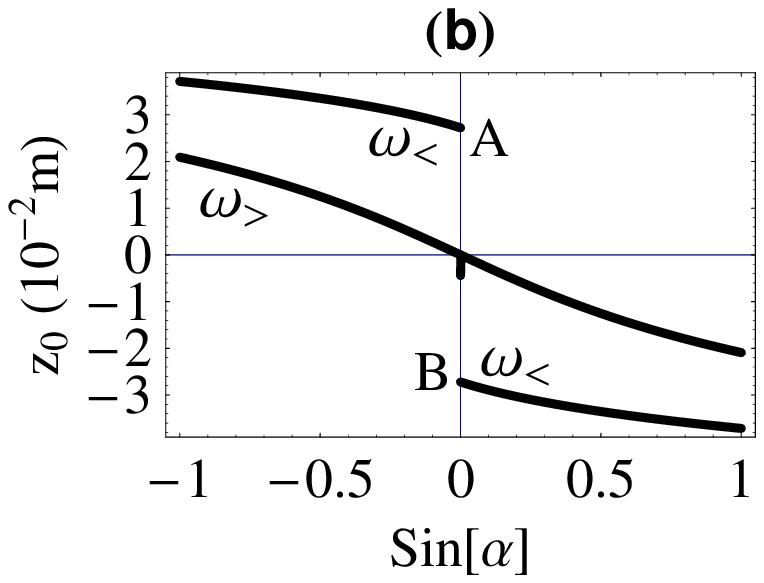}
\caption{{\footnotesize Projections of phase diagram in the (a) $z_{0}-\omega$ and (b) $z_{0}-\text{Sin}\alpha$ planes. The "iso-velocities" in (b) show a first-order transition for $\omega<\omega_{c}$, where $z_{0}$ jumps discontinuously between A and B.}}
\label{Phases}
\end{figure}

\vspace{-0.2cm}

where the last term corresponds to the gravitational potential of both
beads. The zero gravitational potential is taken in the symmetrical point $%
z=0.$ It is clear that Eq. (\ref{Non-sym}) does not hold the
symmetry $z\rightarrow -z$ due to the gravitational term, which will favor
negative positions with lower energy. In this case, the
symmetrical plots from Fig. \ref{symm-potential} will suffer slight
deformations, such as is illustrated in Fig. \ref{nonsymm-potential}, where we can see that for $\omega<\omega_{c}$ the minima splits into two different values as local and global minima. This
behavior is observed, for instance, in an atom or molecule that
presents a degenerated spectrum of energy due to the spherical symmetry. If there are external fields such as electric or
magnetic fields, this spectrum splits into many distinguished levels of
energies (Stark and Zeeman effect).
We can obtain a phase diagram from the minima conditions. When the axis of rotation is in horizontal position, the minima solutions in Eq. (\ref{Z0}) as a function of $\omega$ describes a plot analogous to the 

\onecolumn
\begin{multicols}{2}

\noindent magnetization-temperature phase diagram in a ferromagnet, such as Fig. \ref{Phases}(a) shows. If the z-axis holds different angles $\alpha$ with respect to the horizontal plane, the minimum condition from the effective potential in Eq. (\ref{Non-sym}) describes a diagram as a function of $\text{Sin}\alpha$ similar to the isotherms in a magnetization-magnetic field phase diagram (see Fig. \ref{Phases}(b)). When $\omega<\omega_{c}$, Fig. \ref{Phases}(b) illustrates how the global minimum jumps discontinuously from $-z_{0}$ to $z_{0}$ when $\alpha$ passes through zero, like a first-order transition.            

\vspace{-0.4cm}

\section{Critical Exponents\label{critical}}

Above, we identified the equilibrium position $z_{0}$ as the parameter of
order which measures the degree of breaking of the symmetry. The behavior of
second-order transition suggests us a critical
behavior near the transition point. The effective potential from Eq. (%
\ref{V-efec}) can be written as

\vspace{-0.3cm}

{\small \begin{multline}
V_{ef}(z)=2k\left( R^{2}+r_{0}^{2}\right)-m\omega ^{2}R^{2}+\left( m\omega
^{2}-2k\right) z^{2}\\
-4kr_{0}R\sqrt{1-\left( \frac{z}{R}\right) ^{2}}.
\label{V-efec2}
\end{multline}}

Since nearly below the critical point $\omega _{c}^{2}=\frac{2k(R-r_{0})}{mR}
$ the minimum at $z_{0}=0$ suffers small deviation $z\approx \pm \delta z$,
we can expand the root term in (\ref{V-efec2}) as $\sqrt{1-\left( \frac{z}{R}%
\right) ^{2}}=1-\frac{1}{2}\left( \frac{z}{R}\right) ^{2}-\frac{1}{8}\left( 
\frac{z}{R}\right) ^{4}-\cdot \cdot \cdot $, so that the effective potential
takes the form of a Landau-type expansion near the critical point 

\vspace{-0.3cm}
{\small \begin{equation}
V_{ef}(z_{0})=a_{0}+a_{2}z^{2}+a_{4}z^{4}+\cdot \cdot \cdot ,  \label{Landau}
\end{equation}}

\vspace{-0.3cm}

where the coefficients up to fourth order are

\vspace{-0.3cm}

{\small \begin{subequations}
\begin{eqnarray}
a_{0} &=&-\left[ \omega ^{2}-\left( \frac{R-r_{0}}{R}\right) \omega _{c}^{2}%
\right] mR^{2}  \label{coef-a} \\
a_{2} &=&m\left( \omega ^{2}-\omega _{c}^{2}\right)  \label{coef-b} \\
a_{4} &=&\frac{kr_{0}}{2R^{3}},  \label{coef-c}
\end{eqnarray}
\end{subequations}}

\vspace{-0.3cm}

Furthermore, near $\omega _{c}$ the coefficient in (\ref{coef-b}) becomes

\vspace{-0.2cm}

{\small \begin{equation}
a_{2}\approx 2m\omega_{c}\left(\omega -\omega _{c}\right)=a_{2}^{0}\left(
\omega -\omega _{c}\right) .  \label{approch}
\end{equation}}

\vspace{-0.3cm}

The Landau theory \cite{Landau} is formulated in the framework of the
mean-field approximation, which in our case takes the form of a minimum
condition. Thus, the potential in (\ref{Landau}) up to fourth order must
accomplish the condition

{\small \begin{equation}
\frac{dV_{ef}(z_{0})}{dz}=z_{0}\left[ 2a_{2}^{0}\left( \omega -\omega
_{c}\right) +4a_{4}z_{0}^{2}\right] =0.  \label{minima}
\end{equation}}

For $\omega >\omega _{c}$ the only real solution is $z_{0}=0$, while for $%
\omega <\omega _{c}$ two real minima solutions are obtained

{\small \begin{equation}
z_{0}=\pm \left[ \frac{a_{2}^{0}}{2a_{4}}\left( \omega _{c}-\omega \right) %
\right] ^{1/2}.  \label{minima2}
\end{equation}}

Thus, the parameter of order spontaneously becomes nonzero and grows as $%
\sqrt{\omega _{c}-\omega }$ for angular velocities nearly below $\omega
_{c}. $ The critical exponent $\beta =\frac{1}{2}$ associated with the order
parameter is identical to those obtained by the classical Landau theory for
thermodynamic parameters \cite{Landau, Huang}, such as the magnetic moment in
a ferromagnetic substance, the difference in Zn-Cu occupation in a binary
alloy, the molar volume liquid-gas in a fluid, among other second-order
transitions.

This system could be thought as a mechanical equivalent of a thermodynamic
system, where the state is characterized by three "thermodynamic" variables:
the equilibrium position $z_{0}$ (parameter of order), the angular
velocity $\omega $ (temperature-like variable) and the gravitational force
(pressure-like variable). A change of state can take place if we introduce a "heat reservoir", which in our case is a power source (for example an electric motor) that provides the rotational energy to the hoop. If we consider the complete potential including
the gravitational coupling $W=2mg\text{Sin}\alpha$, the condition in Eq. (\ref{minima})
takes the form

\vspace{-0.3cm}

{\small \begin{equation}
\frac{dV_{ef}(z_{0})}{dz}=2a_{2}^{0}\left( \omega -\omega _{c}\right)
z_{0}+4a_{4}z_{0}^{3}+W=0.  \label{condition2}
\end{equation}}

\vspace{-0.2cm}
  
\end{multicols}

\begin{center}
\begin{table}[tbph]
\begin{tabular}{c|c|c}
\hline\hline
& {\small BHS-System} & {\small Ferromagnet (Ising model)} \\ \hline
{\small Broken Symmetry} & {\small Mirror reflection $z\rightarrow -z$} & {\small Spin reflection $%
s\rightarrow -s$} \\ \hline
{\small Critical Point} & {\small $\omega _{c}=\mid z_{0(\omega \to 0)}\mid \sqrt{\frac{2k%
}{mR(R+r_{0})}}$} & {\small $T_{c}=\mid M_{(T\to 0)}\mid \frac{q\mu _{0}}{k_{B}}$} \\ 
\hline
\begin{tabular}{l}
{\small Parameter of Order} \\ 
({\small near critical point})%
\end{tabular}
& {\small $z_{0}=\left\{ 
\begin{array}{c}
0,\qquad \qquad \qquad \qquad \quad \qquad \ \ \ (\omega >\omega _{c}) \\ 
\pm z_{0(\omega\to 0)}\sqrt{\frac{4R^{2}}{r_{0}(R+r_{0})}}\left[ \frac{\omega
_{c}-\omega }{\omega _{c}}\right] ^{1/2},(\omega <\omega _{c})%
\end{array}%
\right. $} & {\small $M=\left\{ 
\begin{array}{c}
0,\qquad \qquad \qquad \qquad \ \ \ (T>T_{c}) \\ 
\pm M_{(T\to 0)}\sqrt{3}\left[ \frac{T_{c}-T}{T_{c}}\right] ^{1/2},\text{ }%
(T<T_{c})%
\end{array}%
\right. $} \\ \hline
{\small Conjugate Field} & {\small Gravitational Field: $2mg\text{Sin}\alpha $} & 
{\small Magnetic Field $H$} \\ \hline
{\small Susceptibility} & {\small $\chi _{mech}=\left\{ 
\begin{array}{c}
\frac{-R}{8k(R-r_{0})}\left( \frac{\omega -\omega _{c}}{\omega _{c}}\right)
^{-1},\quad (\omega >\omega _{c}) \\ 
\frac{-R}{16k(R-r_{0})}\left( \frac{\omega _{c}-\omega }{\omega _{c}}\right)
^{-1},\quad (\omega <\omega _{c})%
\end{array}%
\right. $} & {\small $\chi _{magn}=\left\{ 
\begin{array}{c}
\frac{1}{q}\left( \frac{T-T_{c}}{T_{c}}\right) ^{-1},\qquad (T>T_{c}) \\ 
\frac{1}{2q}\left( \frac{T_{c}-T}{T_{c}}\right) ^{-1},\qquad (T<T_{c})%
\end{array}%
\right. $} \\ \hline\hline
\end{tabular}%
\caption{{\protect\small Comparison between mechanical and ferromagnetic
systems. $M$ is the magnetic moment, $k_{B}$ the Boltzmann's constant, $%
\protect\mu _{0}$ the Bohr magneton and $q$ the coupling constant of a mean field (which is the magnetic equivalent to the spring constant $k$).
}}
\label{tab:comparison}
\end{table}
\end{center}

\twocolumn

\noindent The mechanical susceptibility, which is understood as the linear response of the
parameter of order (minimum position $z_{0}$) due to an infinitesimal
conjugate field ($W$) is $\chi _{mech}=\frac{\partial z_{0}}{%
\partial W}.$ This can be calculated if we differentiate the Eq. (\ref%
{condition2}) with respect to $W$, obtaining

{\small \begin{multline}
\chi _{mech}=\frac{-1}{2a_{2}^{0}\left( \omega -\omega _{c}\right)
+12a_{4}z_{0}^{2}}\\
=\left\{ 
\begin{array}{c}
\frac{-1}{2a_{2}^{0}}\left( \omega -\omega _{c}\right) ^{-1},\quad \text{
for }\omega >\omega _{c} \\ 
\frac{-1}{4a_{2}^{0}}\left( \omega _{c}-\omega \right) ^{-1},\quad \text{
for }\omega <\omega _{c}%
\end{array}%
\right. .  \label{suscept}
\end{multline}}

from which we deduce the critical exponent $\gamma =1.$ Table \ref{tab:comparison} lists several corresponding features between the mechanical and the ferromagnetic system in the framework of the two-dimensional Ising model under the Mean-Field approximation \cite{Ex}. The results are listed in terms of the equilibrium positions at $\omega=0$ ($z_{0(\omega\to0)}$) and the magnetization at $T=0$ ($M_{(T\to0)}$).

\begin{figure}[t]
\hspace{-1cm}\includegraphics[scale=0.65]{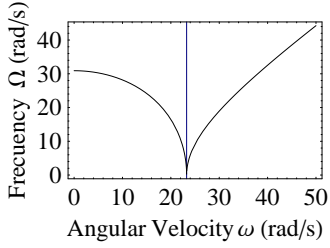}
\caption{{\footnotesize Frecuency of
oscillation as function of $\protect\omega .$ At $\protect\omega _{c},$ the
frecuency becomes zero. Above $\protect\omega _{c}$ the frecuency $\Omega
_{>}$ grows with $\protect\omega $ without limit, while below, the frecuency $\Omega _{<}$ grows as $\protect\omega $
decreases}}
\label{z-function}
\end{figure}

\vspace{-0.3cm}

\section{Small Oscillations\label{Oscillations}}
\vspace{-0.1cm}

We can study small oscillations about the equilibrium positions. For the
symmetrical phase with $\omega >\omega _{c},$ the effective potential in (%
\ref{V-efec}) can be expanded in a Taylor's series about the minimum at $z=0$, from where we obtain an harmonic potential given by 

\vspace{-0.5cm}

{\small \begin{equation}
V_{ef}(\delta z)\approx \text{cons.}+2kr_{0}\left( \frac{1}{R}-\frac{1}{%
\xi }\right) \left( \delta z\right) ^{2}=\text{cons.}+\frac{k_{ef}}{2}%
\left( \delta z\right) ^{2},  \label{harmonic1}
\end{equation}}

\vspace{-0.3cm}

where we can identify an effective spring constant $k_{ef}=4kr_{0}\left( 
\frac{1}{R}-\frac{1}{\xi }\right) .$ We can even identify the effective
mass
defined by the Eq. (\ref{L-gener}) at $z=0$ as $\mu =2m$. Then, the
frequency of oscillation above the critical point is $\Omega _{>}=\sqrt{%
\frac{k_{ef}}{\mu }}=$ $\sqrt{\frac{2kr_{0}}{m}\left( \frac{1}{R}-\frac{1}{%
\xi }\right) },$ where $\left\vert \xi \right\vert >\left\vert R\right\vert $. The frecuency $\Omega _{>}$
approaches zero near $\omega _{c}$, and the period of oscillation becomes
infinite in the threshold of stability at $z=0$ (see Fig. \ref{z-function}). This frecuency increases
without limit with $\omega .$ For the breaking phase with $\omega <\omega
_{c}$, we can perform a similar analysis about the minima at $z=\pm \sqrt{%
R^{2}-\xi ^{2}}.$ In this case, the effective constant is $%
k_{ef}=4kr_{0}\left( \frac{R^{2}}{\xi ^{3}}-\frac{1}{\xi }\right) $ and the
effective mass is $\mu =2m\frac{R^{2}}{\xi ^{2}}$. Hence, the frequency is $%
\Omega _{<}=$ $\sqrt{\frac{2kr_{0}}{m}\left( \frac{1}{\xi }-\frac{\xi }{R^{2}%
}\right) }$ where $\left\vert \xi \right\vert <\left\vert R\right\vert .$ Here, as $\omega $ decreases, the function $\xi (\omega )$ decreases and
reaches its minimum value at $\xi =r_{0}$ when the hoop comes to a complete
rest with $\omega =0$. As the
angular velocity decreases, the frecuency $\Omega _{<}$ below the critical
value becomes greater and reaches its maximum value $\Omega _{<}=\sqrt{\frac{%
2k}{m}\left( 1-\left( \frac{r_{0}}{R}\right) ^{2}\right) }$ when $\omega =0$.

\vspace{-0.3cm}

\section{Chaotic Behavior\label{chaos}}

\begin{figure}[t]
\hspace{-2.3cm}\includegraphics[scale=0.53]{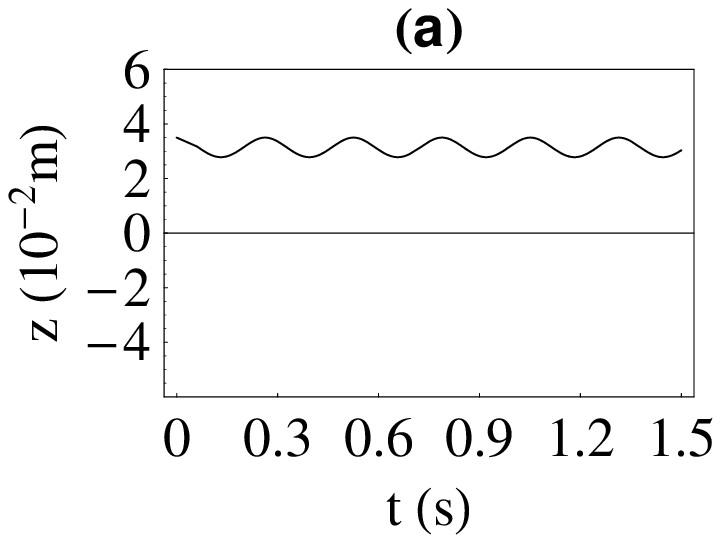} \hspace{-2.5cm}
\includegraphics[scale=0.53]{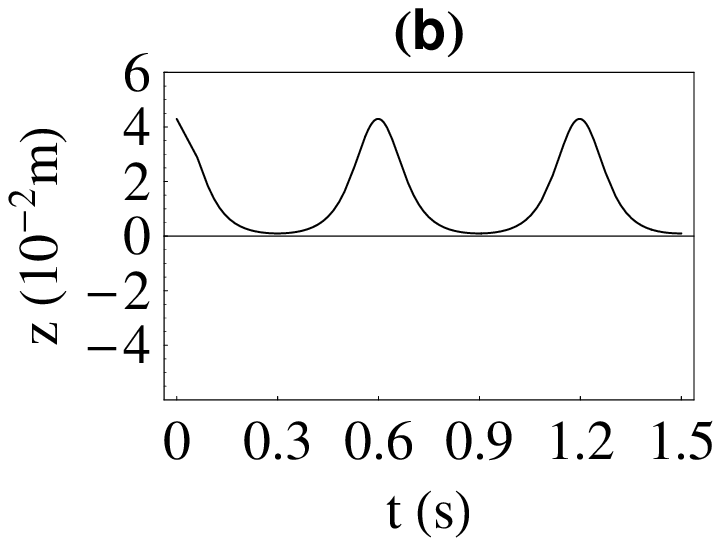}

\vspace{-0.2cm}

\hspace{-2.3cm}
\includegraphics[scale=0.53]{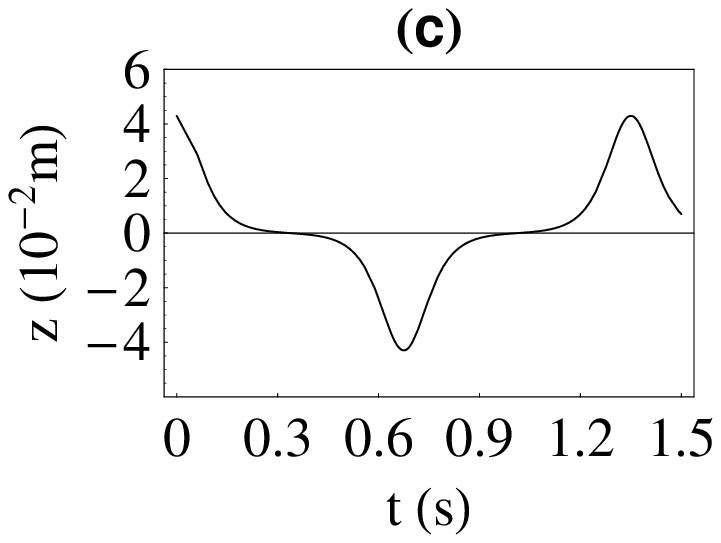} \hspace{-2.5cm}
\includegraphics[scale=0.53]{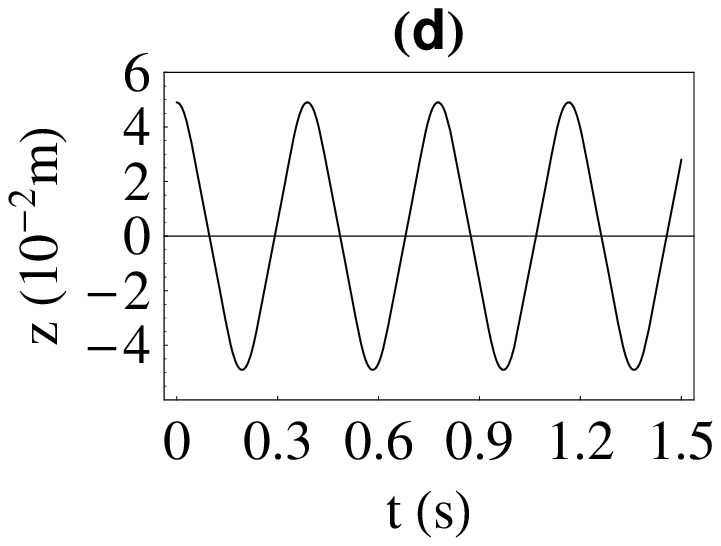}

\vspace{-0.3cm}
\hspace{-1cm}
\includegraphics[scale=0.6]{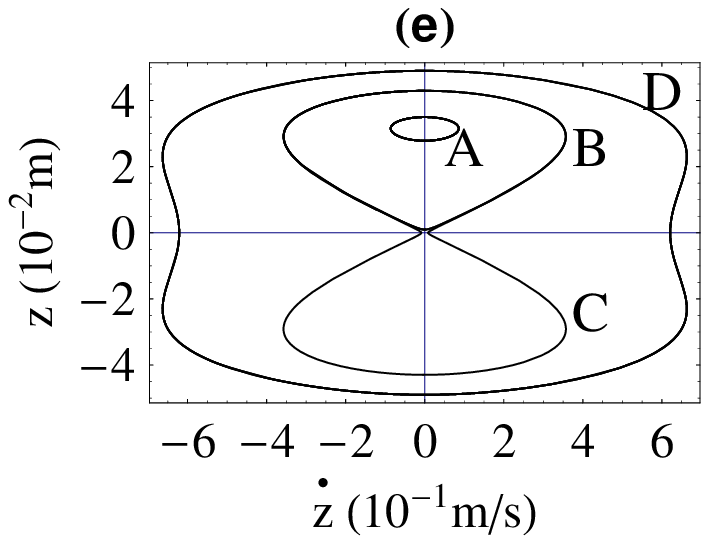}
\caption{{\footnotesize Graphical solutions for $\protect\omega<\protect\omega _{c}$.
The position as a function of time is shown for (a) small perturbations at $%
z>0$, greater perturbations just (b) below and (c) above the jump over $z=0$%
, and (d) for enormous perturbations. Each solution (A,B,C and D
respectively) is shown (e) in the phase space.}}
\label{Chaos-solut1}
\end{figure}

The complete description of the system is obtained as solution from the
Lagrange's equations associated with the generalized coordinate $z$

\vspace{-0.2cm}

{\small \begin{equation}
\frac{d}{dt}\left( \frac{\partial L}{\partial \overset{\cdot }{z}}\right) -%
\frac{\partial L}{\partial z}=0.  \label{Lagrange}
\end{equation}}

With the lagrangian defined by Eq. (\ref{L-gener}) and the effective
potential from Eq. (\ref{V-efec}), we obtain the equation of motion

\vspace{-0.5cm}

{\small \begin{multline}
\overset{\cdot \cdot }{z}+\left( \frac{z}{R^{2}-z^{2}}\right) \overset{\cdot 
}{z}^{2}+\left( \frac{2k}{mR^{2}}-\frac{\omega ^{2}}{R^{2}}\right) z^{3}+%
\frac{2kr_{0}}{mR^{2}}z\sqrt{R^{2}-z^{2}}\\
-\left( \frac{%
2k}{m}-\omega ^{2}\right)z=0,  \label{motion}
\end{multline}}

\vspace{-0.3cm}

which is a quite complex equation to solve. We note that this equation
involves non-linear terms, which suggests us a chaotic behavior. From data given in Sec. \ref{Experimental}, we consider numerical solutions with different initial positions
around the equilibrium points (the entrance $\overset{\cdot 
}{z}(0)$ was always taken zero). 
Figs. in \ref{Chaos-solut1} display the behavior of the position $z$ as
function of time and in the phase space $\left( z\text{ }Vs.\text{ }\overset{%
\cdot }{z}\right) $ for $\omega <\omega _{c}$. We see clearly that for small
perturbations, the oscillations are restricted in regions with $z>0.$ In
the phase space, this motion describes elliptic-like paths (we also see that
these paths are not perfect ellipses due to anharmonic terms). If we
increase the perturbation, the system reaches the transition point, where the
spring just oversteps the local maximum at $z=0$, and the oscillations extend
to both regions. For greater perturbations, the motion increases around all
the hoop.

\begin{figure}[t]
\hspace{-1.7cm}\includegraphics[scale=0.55]{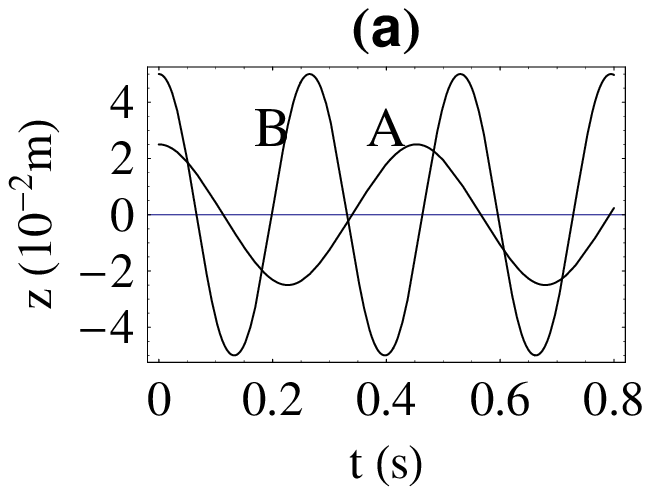} \hspace{-2.5cm}
\includegraphics[scale=0.55]{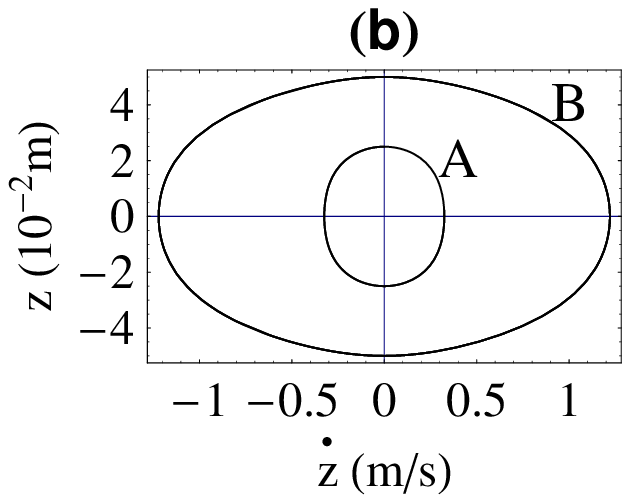}
\caption{{\footnotesize Graphical solutions for $\protect\omega>\protect\omega_{c}$. The
position as a function of time is shown in (a) for (A) small perturbations and
(B) greater perturbations. Each solution (A and B respectively) is
shown (b) in the phase space.}}
\label{Chaos-solut2}
\end{figure}

\begin{figure}[t]
\hspace{-2.3cm}\includegraphics[scale=0.51]{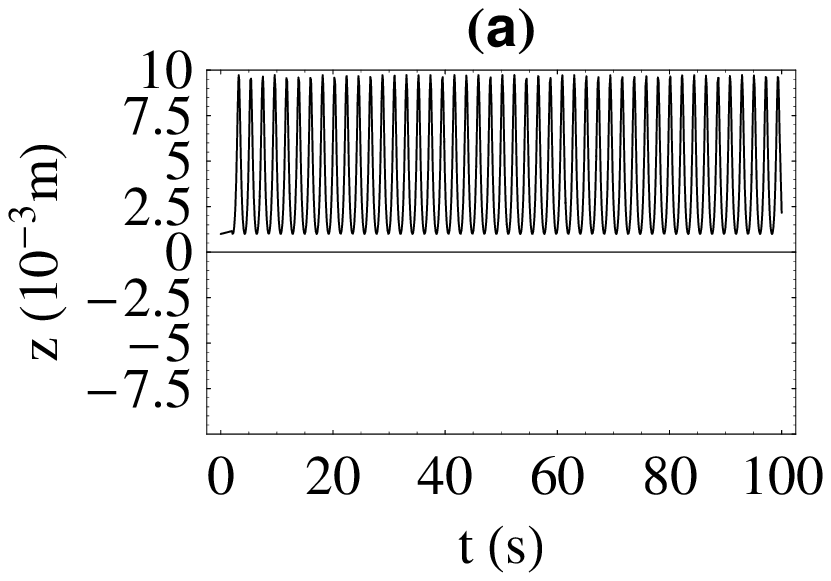} \hspace{-2.5cm}
\includegraphics[scale=0.51]{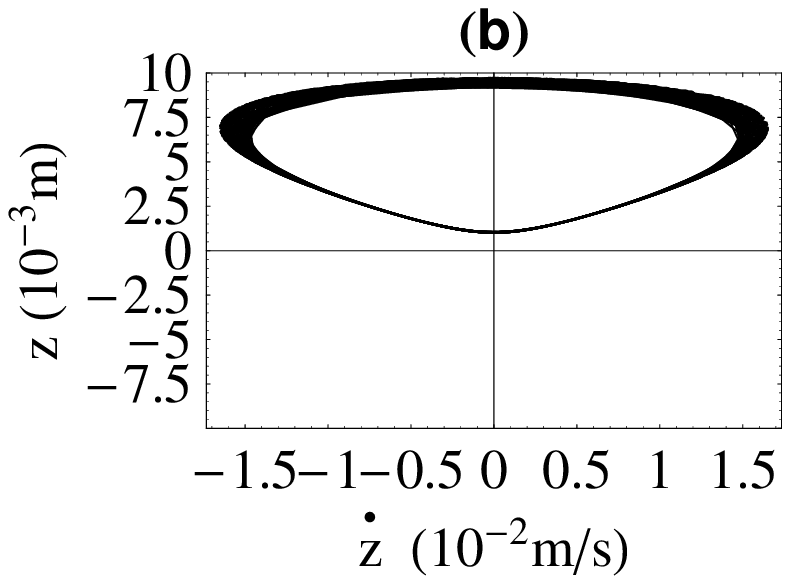}

\hspace{-2.3cm} 
\includegraphics[scale=0.51]{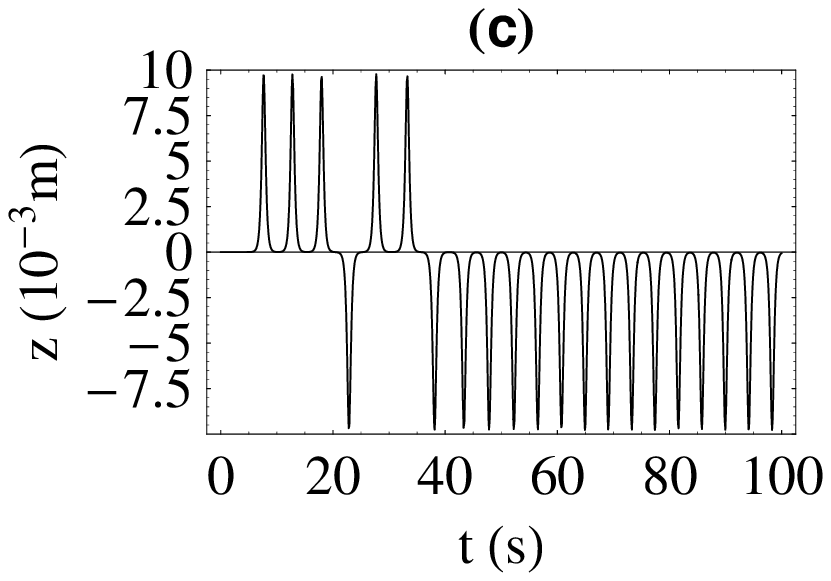} \hspace{-2.5cm}
\includegraphics[scale=0.51]{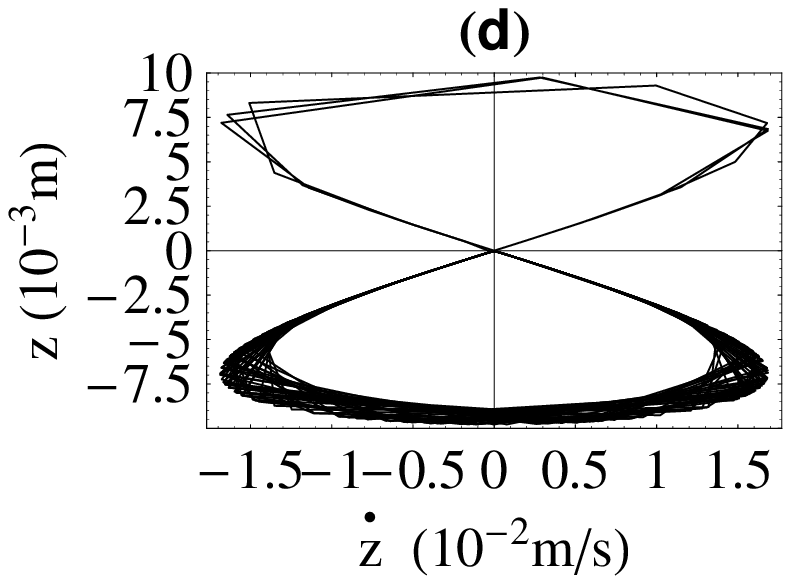}

\hspace{-2.3cm}
\includegraphics[scale=0.51]{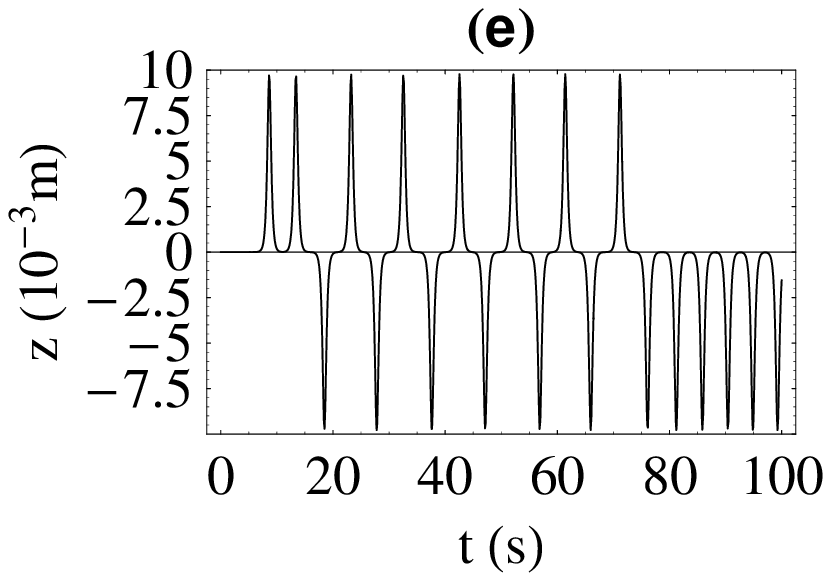} \hspace{-2.5cm}
\includegraphics[scale=0.51]{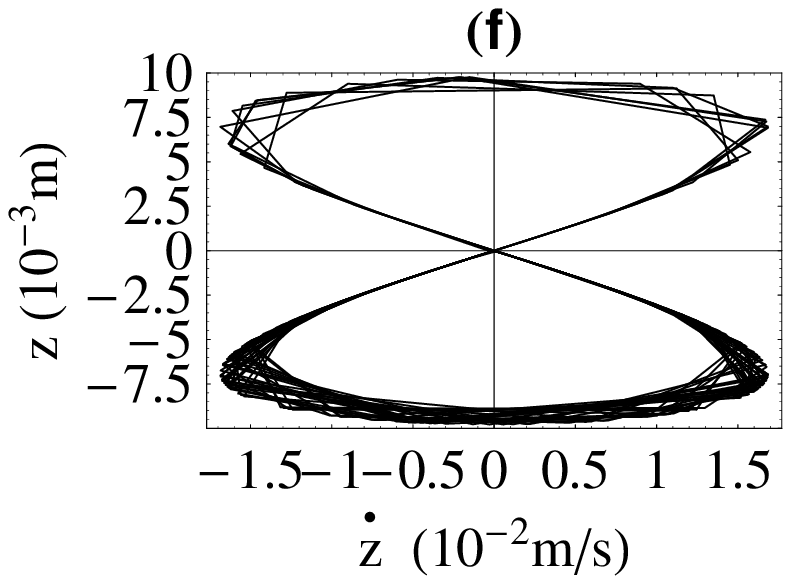}
\caption{{\footnotesize Graphical solutions for $\protect\omega$ slightly below $\protect%
\omega_{c}$. The position as a function of time is shown for (a) very small
perturbations at $z>0$ and (c),(e) greater perturbations. Each solution
((b),(d) and (f) respectively) is shown in the phase space. These plots
exhibit a chaotic behavior.}}
\label{Chaos-solut3}
\end{figure}

\begin{figure}[h] 
\hspace{-2cm}\includegraphics[scale=0.59]{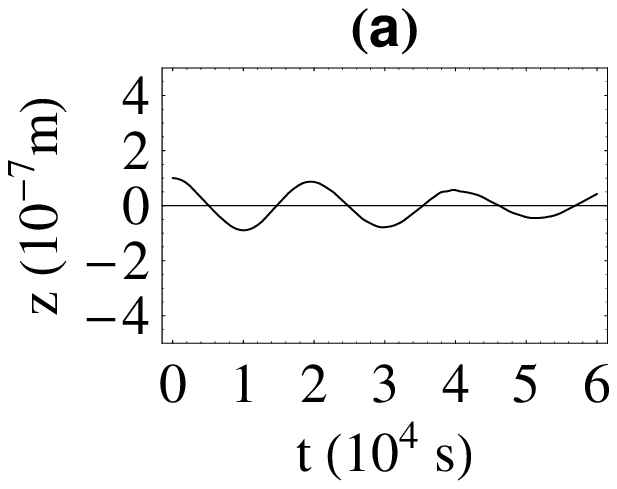} \hspace{-2.7cm}
\includegraphics[scale=0.52]{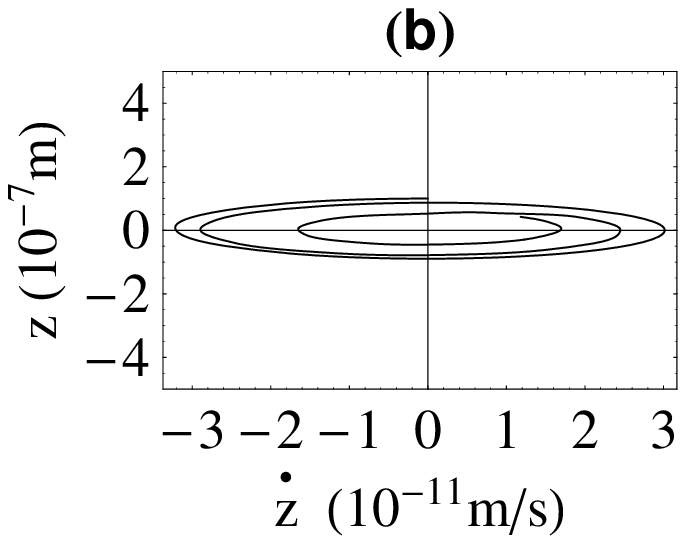}

\hspace{-2cm}
\includegraphics[scale=0.51]{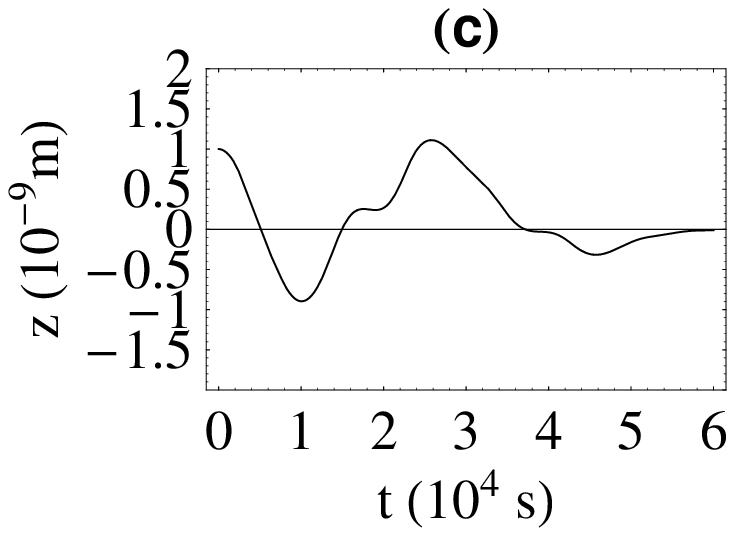} \hspace{-2.3cm}
\includegraphics[scale=0.52]{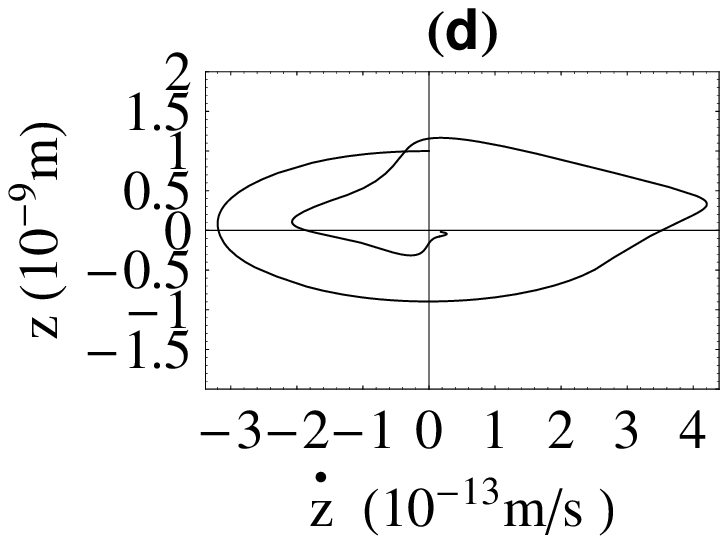}
\caption{{\footnotesize Graphical solutions for $\protect\omega$ = $\protect\omega_{c}$. The
position as a function of time is shown for (a),(c) small perturbations.
Each solution ((b) and (d) respectively) is shown in the phase space. These
plots exhibit a long oscillation with chaotic behavior.}}
\label{Chaos-solut4}
\end{figure}

On the other hand, we can see in Fig. \ref{Chaos-solut2} that for $%
\omega >\omega _{c}$, the motion is developed around the equilibrium point at 
$z=0$, in agreement with the results of Sec. \ref{Oscillations}. Figs.
in \ref{Chaos-solut3} represent solutions for a value below but near the
critical value, where we introduced small perturbations. For a slight perturbation
($\delta z_{0}\sim5.92\times 10^{-3} m$), the system
develops a motion in the breaking region. But if this perturbation
is changed into $\delta z_{0}\sim6.926\times 10^{-3} m$, the system presents random oscillations between both regions,
where the graphs in the phase space describes chaotic paths. The pattern of this chaotic behavior is in extrem sensitive to very tiny changes of initial conditions (with differences of just $\sim1\times 10^{-13} m$). Exactly on the
critical value (Fig. \ref{Chaos-solut4}), the system presents very long
and irregular oscillations.

\vspace{-0.3cm}

\section{Experimental Description\label{Experimental}}

We made a demostrative BHS-apparatus using a hoop with radius $R=6$ $cm$, a
soft spring with natural length $r_{0}=4.6$ $cm$ and constant $k=8.81$ $N/m$%
, and two iron rings used as beads, each one with mass $m=7.6$ $g$. The hoop
was coupled to the axis of a DC motor with adjustable velocity (which will work like a "heat reservoir").

\begin{figure}[t]
\centering\includegraphics[scale=0.6]{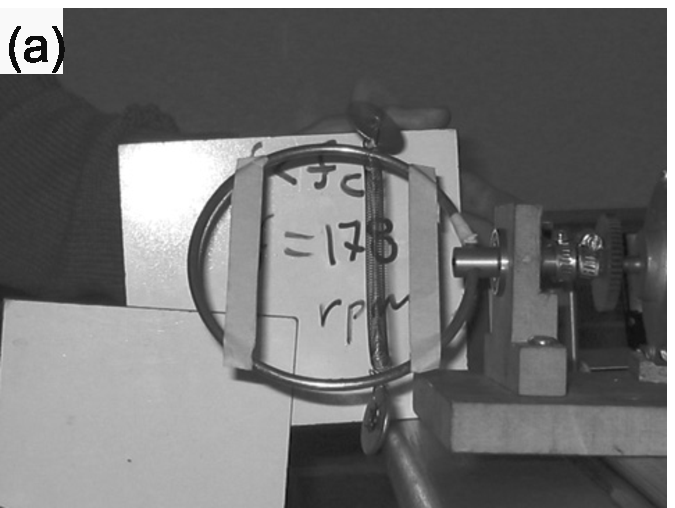}
\includegraphics[scale=0.6]{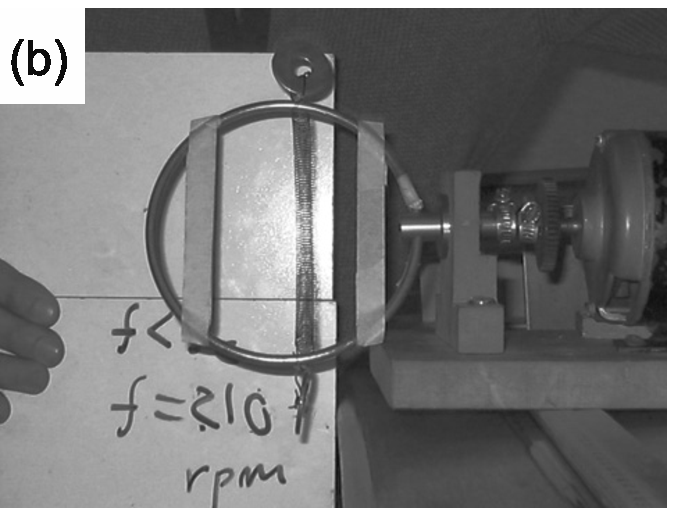}
\includegraphics[scale=0.6]{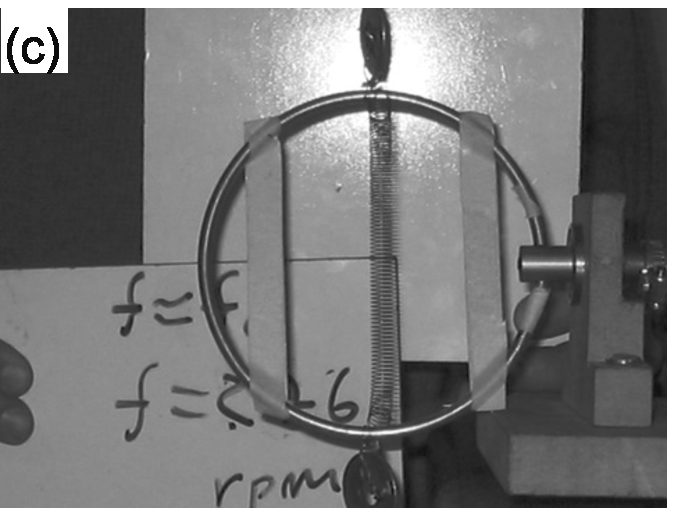}
\includegraphics[scale=0.6]{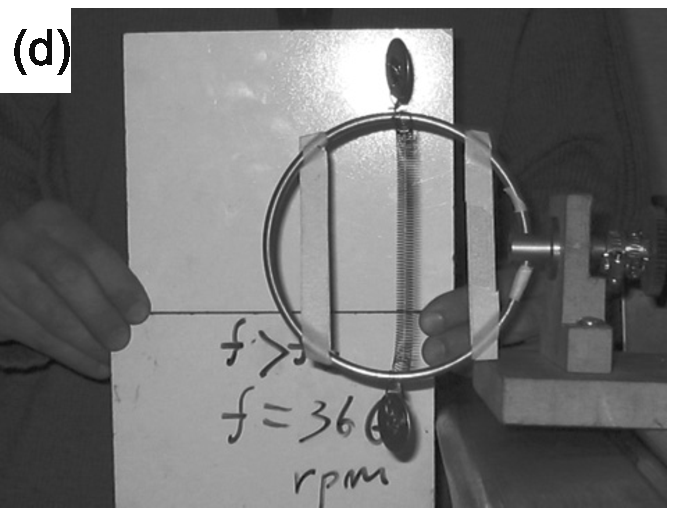}
\caption{{\footnotesize Experimental device for frecuency of rotation (a) $f=178RPM$ ($\protect%
\omega =18.64rad/s$), (b) $f=210RPM$ ($\protect\omega =21.99rad/s$), (c) $%
f=246RPM$ ($\protect\omega =25.13rad/s\approx \protect\omega _{c}$) and (d) $%
f=366RPM$ ($\protect\omega =38.32rad/s>\protect\omega _{c}$).}}
\label{photos}
\end{figure}

Aligning the
rotational axis in a horizontal position, the system describes the two regimes
of symmetry. The experimental observation is illustrated in Fig. \ref{photos}
according to the angular velocity. The measured critical velocity was $\omega _{c}=25.24\pm 2.1$ $%
rad/s$ ($241\pm 20$ $RPM$), which is in a quite good agreement with the results from table \ref%
{tab:stability}, where a critical point at $\omega _{c}=23.26$ $rad/s$ ($%
222.1$ $RPM$) is expected. In addition to the motion at $\omega_{c}$, Fig. \ref{photos}
shows two cases with $\omega <\omega _{c}$ and one case with $\omega >\omega
_{c}.$ When the velocity was decreased below the critical value, we noted that
the chosen direction ($+z$ or $-z$) was highly sensitive to the way that the
velocity was reduced and to the mechanical fluctuations near the transition. When the
velocity was slowly reduced in a quasi-static process across the transition,
the equilibrium position was maintained at $z=0$ (although we lubricate the
hoop, the unstable equilibrium found some stability due to the small
friction), but after some hesitating seconds
the system finally fell into one of the stable minima (long relaxation
time), breaking the symmetry. However after several trials, we could see that
the symmetry was hidden in the sense that we obtained about the same number
of breakings into $+z$ as into $-z$, i.e. both posibilities have the same
probability of occurrence. Thus, we can see how the symmetry breaking is
spontaneous, where the non-symmetrical states present two equally probable
possibilities with the same energy.
On the other hand, the device had a high sensitivity to the precision of the
horizontal alignment. When the rotational axis was slightly deviated with $\omega>\omega_{c}$, the
system always shows a breaking in one direction due to the gravitational
potential, in agreement with the behavior described by Fig. \ref%
{nonsymm-potential}. We estimated experimental explicit breaking within a range $\alpha<0.1$ $%
rad $ ($\lesssim \pm 5^{\circ }$).

\section{Summary and Conclusions\label{conclusions}}

In this work we studied a bead-hoop-spring (BHS)
system under rotation as was described through the sections. The system
presents different stability situations according to the angular velocity
of the hoop, exhibiting a second order transition behavior, and where the
position $z_{0}$ of the center of mass measured from the center of the hoop
defines a parameter of order. The symmetry transformation $z\rightarrow -z$
of the lagrangian is manifest if the angular velocity is above
a critical value $\omega _{c}$, where the
equilibrium position at $z_{0}=0$ is stable. Below $\omega _{c}$, this
equilibium becomes unstable, and appears two degenerated minima with $z_{0}\neq0$.
Thus, with only internal interactions (spring force and normal force between hoop and beads), the system exhibits an SSB below the critical value. If the
gravitational potential is considered, the stability regions suffer a
deformation, changing the equilibrium position at $z=0$ into
another different from zero, and spliting the minima of energy at $z_{0}\neq0$ into two non-equivalent minima.
Under these circumstances, the system
shows a preferential direction exhibiting explicit symmetry breaking and a first-order transition behavior. The experimental observations were in quite good agreement with the theoretical description, where the critical velocity was measured within an $8\%$ of precision.

The system belongs to the same universality class that many thermodynamic
systems, exhibiting a Landau expansion. From a mean-field-like
approximation, we obtained the critical exponent $\beta =\frac{1}{2}$ and $%
\gamma =1$ associated with the parameter of order and the
"mechanical susceptibility" respectively. 
       
The frecuency of small oscillations about the stable minima was also
calculated. Above $\omega _{c}$, the frecuency $\Omega _{>}$ increases without limit with $\omega .$ Near the critical
value, the motion about $z=0$ describes long-period oscillations, which becomes into unbounded motion when $\omega$ passes through the critical point. Below $\omega _{c}$, the
frecuency $\Omega _{<}$ about the new minima increases with $\omega$ decreases.  

The equation of motion exhibits some chaotic behavior near the critical
value, where the system reacts in a random way for similar perturbation
entrance.

This analogy between the mechanical system and a thermodynamic system is
highly suggestive, and it is very tempter to apply a deeper study of
fluctuations below the critical point through the definition of a
correlation function and the implementation of a fluctuation-dissipation
theorem \cite{Landau, Huang}, which could provide some insights about the
physical sense of these concepts.

\vspace{-0.3cm}

\section*{Acknowledgment}

We thank Rodolfo Diaz and William Herrera from Universidad Nacional de Colombia for their comments, which improve the content of this paper. We also thank to Mr. Francisco and Yuri from Escuela Colombiana de Ingenier\'{\i}a for their collaboration in the development of the experimental device.

\vspace{-1cm}

\renewcommand{\refname}{}


\end{document}